\documentclass[]{spie}  

\usepackage{amsmath,amsfonts,amssymb}
\usepackage{graphicx}
\usepackage[colorlinks=true, allcolors=blue]{hyperref}

\title{Developing post-coronagraphic, high-resolution spectroscopy for terrestrial planet characterization on ELTs}

\author[a,b]{N. Jovanovic}
\author[a,c,d,e]{O. Guyon}
\author[f]{T. Kotani}
\author[g,h]{H. Kawahara}
\author[f]{K. Hosokawa}
\author[a]{J. Lozi}
\author[c]{J. Males}
\author[i]{M. Ireland}
\author[e,f,j]{M. Tamura}
\author[k]{D. Mawet}
\author[b,l]{C. Schwab}
\author[l,m]{B. Norris}
\author[m]{S. Leon-Saval}
\author[m]{C. Betters}
\author[m]{P. Tuthill}

\affil[a]{National Astronomical Observatory of Japan, Subaru Telescope, Hilo, HI, 96720, U.S.A.}
\affil[b]{Department of Physics and Astronomy, Macquarie University, NSW 2109, Australia}
\affil[c]{Steward Observatory, University of Arizona, Tucson, AZ, 85721, U.S.A.}
\affil[d]{College of Optical Sciences, University of Arizona, Tucson, AZ 85721, U.S.A.}
\affil[e]{Astrobiology Center of NINS, 2-21-1, Osawa, Mitaka, Tokyo, 181-8588, Japan}
\affil[f]{National Astronomical Observatory of Japan, 2-21-1 Osawa, Mitaka, Japan}
\affil[g]{Department of Earth and Planetary Science, The University
of Tokyo, Tokyo 113-0033, Japan}
\affil[h]{Research Center for the Early Universe, School of Science,
The University of Tokyo, Japan}
\affil[i]{Research School of Astronomy \& Astrophysics, Australian National University, Canberra ACT 2611, Australia}
\affil[j]{The University of Tokyo, Tokyo 113-0033, Japan}
\affil[k]{California Institute of Technology, Pasadena, CA, 91125, U.S.A.}
\affil[l]{Australian Astronomical Observatory, 105 Delhi Rd, North Ryde NSW 2113, Australia}
\affil[m]{Sydney Institute for Astronomy (SIfA), Institute for Photonics and Optical Science (IPOS), Sydney Astrophotoinc Instrumentation Laboratory (SAIL), School of Physics, University of Sydney, NSW 2006, Australia}

\authorinfo{Nemanja Jovanovic: E-mail: jovanovic.nem@gmail.com, Telephone: 1 626 395 1214}

\begin{document} 
\maketitle

\begin{abstract}
Spectroscopic observations are extremely important for determining the composition, structure, and surface gravity of exoplanetary atmospheres. High resolution spectroscopy of the planet itself has only been demonstrated a handful of times. By using advanced high contrast imagers, it is possible to conduct high resolution spectroscopy on imageable exoplanets, after the star light is first suppressed with an advanced coronagraph. Because the planet is spatially separated in the focal plane, a single mode fiber could be used to collect the light from the planet alone, reducing the photon noise by orders of magnitude. In addition, speckle control applied to the location where an exoplanet is known to exist, can be used to preferentially reject the stellar flux from the fiber further.

In this paper we will present the plans for conducting high resolution spectroscopic studies of this nature with the combination of SCExAO and IRD in the H-band on the Subaru Telescope. This technique will be critical to the characterization of terrestrial planets on ELTs and future space missions.
\end{abstract}

\keywords{Extreme AO, Adaptive optics, High contrast imaging, Exoplanets, Coronagraphs, Imager, Fiber injection}

\section{INTRODUCTION}
\label{sec:intro}  
The aim of high contrast imaging is to be able to image an extremely faint brown dwarf, disk or exoplanet in close proximity to its host star ($<2$" away). Atmospheric turbulence distorts the incoming wavefronts preventing a diffraction-limited image from being achieved at ground-based observatories. This has a disastrous effect for high contrast imaging as the photons of the bright star are now spread over a much larger area swamping the faint signal from the companion/disk. Adaptive optics systems try to counter act this by restoring light to the central core of the point-spread function (PSF), reducing the photon-noise at the location of the companion. Extreme AO (ExAO) systems push this one step further by achieving Strehl ratios of $90\%$ in H-band in better than median seeing conditions~\cite{Mac2014,Vigan2016}.

Once the light is mostly localized to the core of the PSF, it is common to exploit a coronagraph to try and suppress it. This not only suppresses the on-axis star light but also minimizes diffraction effects downstream reducing the brightness of the halo away from the star where the companion is located. Differential imaging techniques such as angular differential imaging (ADI), which relies on the motion of the companion as the field rotates during an observing sequence (when operating in fixed pupil mode), can be used to discriminate between the speckles and companion and enhance contrast.  However, at small angular separations ($<0.3$") it is still difficult to achieve sufficient contrast to image several Jovian mass planets, let alone terrestrial planets. Indeed $51$ Eri at a modest $2$ Jupiter masses is the lowest mass companion imaged at $<0.5$"(or at any separation for that matter) to date~\cite{mac2015}. The contrast requirements on characterizing planets in this region are even steeper. 

One approach to overcome this difficulty in achieving the contrast required for characterization was proposed by Snellen et al.~\cite{snellen2015}. The concept is to use a fiber to couple the light from the planet to a high-resolution spectrograph. For this to work the position of the planet must be known, so it has to have been previously detected. Ideally, the fiber would be matched to efficiently couple in the light from the planet. In this regime the fiber will also offer some suppression of the stellar flux in the vicinity of the planet. If a single mode fiber is used, then it is possible to use an upstream deformable mirror to modulate the phase of the field at the location of the planet to enhance suppression of the stellar halo while efficiently coupling in the planet light~\cite{mawet2017}. The light can then be dispersed with a high-resolution spectrograph. The signal from the planet will be shifted in frequency with respect to the signal from the star owing to the large relative motion of the two bodies. By cross-correlating template spectra from molecules one would expect to see in the atmosphere of the planet under study, a signal can be extracted from the photon-noise limited data. 

This approach uses multiple tiers to suppress the contamination from the star. It was recently shown through simulation that this technique could enable the characterization of Proxmia Centauri b on an ELT~\cite{wang2017}. In this work this technique was dubbed ``high dispersion coronagraphy " (HDC). It has not been experimentally demonstrated on-sky. Here we outline the ongoing work to enable this mode by connecting the Subaru Coronagraphic Extreme Adaptive Optics (SCExAO) instrument~\cite{Nem2015a} with the IR Doppler (IRD) Instrument~\cite{Kotani2014}. We outline the status of this effort and future work we intend to carry out. The aim of this work is to demonstrate that this technique can be used to bridge the gap between where we are now and where we need to be in regards to contrast required to characterize planets in order to build an optimized terrestrial planet imager for the ELTs.

\section{Implementing High Dispersion Coronagraphy with SCExAO/IRD}
\begin{figure*}
\centering 
\includegraphics[width=0.95\linewidth]{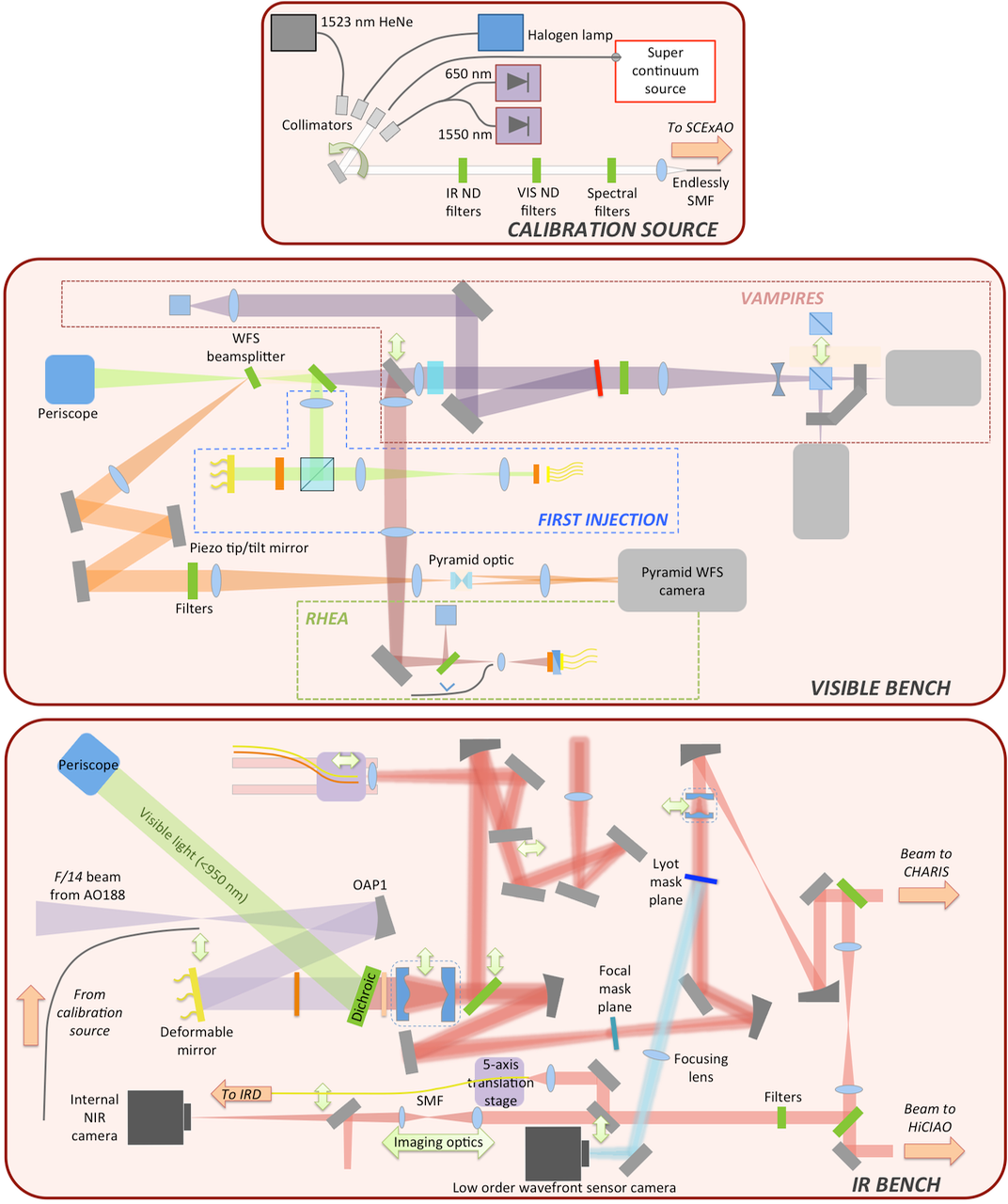}
\caption{\footnotesize Schematic of the SCExAO instrument. (Top) The calibration source which allows the user to select between a super continuum source, 1523 nm HeNe laser, $650/1550$~nm laser diode and a halogen lamp. (Middle) Visible bench of SCExAO which hosts the pyramid wavefront sensor. (Bottom) Infrared bench of SCExAO which hosts the coronargaph, deformable mirror and the fiber injection. Dual head green arrows indicate that a given optic can be translated in/out of or along the beam. Orange arrows indicate light entering or leaving the designated bench at that location. Labels have been suppressed for non-relevant optics.}
\label{fig:schematic}
\end{figure*}

A schematic of the SCExAO instrument as of May $2017$ is shown in Fig.~\ref{fig:schematic}. A comprehensive overview of the instrument can be found in Jovanovic \textit{et. al.} ($2015$), so here we offer a basic overview and focus on the modules that will be key to realizing HDC with SCExAO/IRD. 

\subsection{Wavefront control}
The first stage to improving contrast is to restore the light from the central star to the core of the PSF. This is achieved through two stages of AO correction. The first stage correction is provided by AO$188$~\cite{Min2010}, the facility AO system at Subaru Telescope. AO$188$ offers Strehl ratios of $30-40\%$ in the H-band, in median seeing conditions. The second stage correction is implemented within SCExAO. A $2000$-element deformable mirror (DM, Boston MicroMachines Corporation) is used to drive the correction, which is determined by the pyramid wavefront sensor (PyWFS). The PyWFS, located on the visible bench uses light between $800$-$950$~nm and operates at $2$~kHz on targets with I magnitude $<9$. The first stage correction provided by AO$188$ ensures that the beam delivered to the PyWFS is within the linear range of the sensor. Owing to the PyWFS's exquisite sensitivity, a correction at both higher spatial and temporal frequencies is achieved, which can result in Strehl ratios of up to $90\%$ in H-band, in better than median seeing conditions~\cite{Lozi2016a,Lozi2016b}. 

\subsection{Coronography}
With the light mostly returned to the central core of the PSF, the second stage involves suppressing the starlight with the use of a coronagraph. SCExAO makes use of numerous types of coronagraphs including the Phase Induced Amplitude Apodization (PIAA)~\cite{Guyon2003}, vector vortex~\cite{Mawet2010,Kuhn2016}, eight octant phase mask, classical Lyot and shaped pupil to name a few~\cite{Nem2015a}. These coronagraphs are optimized for different regimes. For example the PIAA and vector vortex offer very small inner working angle ($\sim1~\lambda/D$) but are highly sensitive to the Strehl ratio and tip/tilt jitter. On the other hand the shaped pupil coronagraph has an inner working angle of $5~\lambda/D$ but as it is located in a pupil plane, is insensitive to tip/tilt jitter. 

Another important aspect to consider is achromaticity. The IRD instrument operates across the y-H bands and so a coronagraph that can suppress starlight in the region of a known planet across all bands is ideal. The vortex as implemented on SCExAO can only operate over the H-band currently. The PIAA can operate across all $3$ bands but will have a compromised inner working angle at shorter wavelengths. A classical Lyot will work well but also offer a compromised inner working angle as well. 

Initial experiments will be focused on the H-band only, and on known exoplanets within the control radius of the DM ($0.9$" in H-band) but not closer than $0.3$" where the contrast requirements are a challenge. In this regime any of the coronagraphs mentioned above could be implemented. Eventually the system will be optimized to operate across all bands (y-H).   

\subsection{Low order wavefront sensing}
Despite the fact that the PyWFS restores the PSF in good conditions to a Strehl ratio of $90\%$, the correction is conducted in a non-common path to the coronagraph on the IR bench and at a different wavelength ($800$--$900$~nm versus y-H bands). This creates non-common path and chromatic errors in the correction which result in stellar leakage at the coronagraphic mask. For this reason a low order wavefront sensor (LOWFS) forms the third stage to conducting HDC effectively on SCExAO. 

The light diffracted by a phase-based focal plane mask is reflected at the Lyot stop, and imaged onto a LOWFS camera, depicted in Fig.~\ref{fig:schematic}. The diffracted light from the coronagraphic mask can be used to drive a low order correction at the location of the coronagraph itself. It is administered in the form of an offset sent to the PyWFS, which is the sole driver of the DM. 

The Lyot-based LOWFS has been shown to deliver tip/tilt residuals on-sky better than $10^{-3}~\lambda/D$ with the various phase mask coronagraphs on SCExAO~\cite{singh2017}. This will stabilize the correction provided by the PyWFS and enhance stellar rejection.

\subsection{Fiber injection}
\begin{figure*}
\centering 
\includegraphics[width=0.99\linewidth]{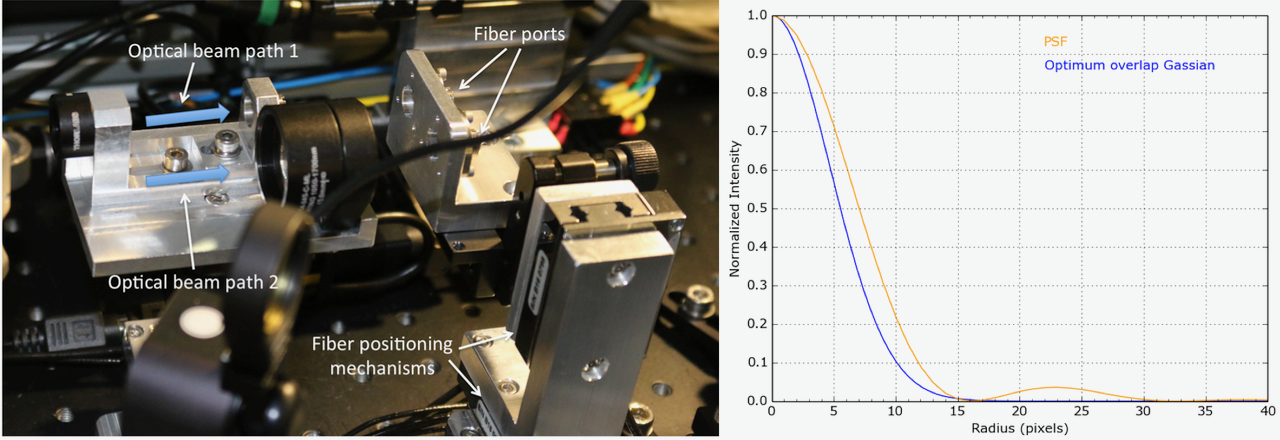}
\caption{\footnotesize (Left) The image shows the fiber injection in SCExAO for high dispersion coronography. There are two optical beam paths for testing both a single-mode and multimode fiber. (Right) The optimum overlap Gaussian for an Airy beam generated at the focus of Subaru Telescope.}
\label{fig:fibinj}
\end{figure*}

The fiber injection is key to positioning the fiber accurately in the focal plane, which will couple the planet light efficiently and aid in filtering out the unwanted starlight. For this reason this forms the fourth stage of the system. The fiber injection is depicted in Fig.~\ref{fig:fibinj} and can also be seen near the LOWFS in Fig.~\ref{fig:schematic} above.  

The fiber injection consists of a $5$ axis stage. Three axes can be driven remotely in X, Y and focus and two axes can only be driven manually (tip/tilt). The linear stages used (Newport, Conex-AG-LS25-27P) allow for incremental fiber motions of $200$~nm step size, and a unidirectional repeatability of $150$~nm. The fiber injection rig can support two fibers at once. To switch between fibers the horizontal stage needs to be translated laterally. 

The system is designed to support two fibers so that a single and mulitmode fiber could be compared. The multimode fiber is easier to couple into offering higher efficiency for the planet at the expense of more stellar leakage. The single mode is superior at filtering the star light but offers reduced throughput for the planet. In either case it is critical to match the size of the core to the size of the planet, $\sim\lambda/D$. Since the core sizes are vastly different, $8~\mu$m for the single mode fiber (SMF28-J9) versus $60~\mu$m for the multimode fiber (OFS F8950), two optical systems with varying focal ratios were needed. 

The right panel of Fig.~\ref{fig:fibinj} shows the theoretical optimum overlap Gaussian profile for the Airy beam generated at the focus of Subaru Telescope (i.e. taking into account the $30\%$ fractional size of the central obstruction, with no wavefront error). Based on the fact that the mode field diameter for the single-mode fiber is fixed at $10.4~\mu$m, the focal ratio needs to be adjusted so that the Airy spot is scaled to preserve the ratio of the curves in Fig.~\ref{fig:fibinj}. A focal ratio of $f/5$ was needed for optimum coupling. Given the $9$~mm beam diameter a single $f=45$~mm lens (Thorlabs, AC254-045-C) was used to create the $f/5$ beam and is indicated as optical beam path $2$ in the left panel of Fig.~\ref{fig:fibinj}. For the multimode fiber, a $f/26$ beam was used. This was realized with the combination of a $f=50$~mm lens (AC127-050-C-ML), and a $f=7.5$~mm (AC050-008-C-ML) and can be identified as optical beam path $1$ in the left panel of Fig.~\ref{fig:fibinj}. These optical systems are co-mounted on a single translation stage which can be used to switch between them efficiently. 

\subsection{The fibers and alignment}
As presented in the early work by Snellen et al.~\cite{snellen2010} it is important to remove the contamination from both the stellar spectrum and the tellurics from Earths atmosphere. This can most easily be done by using a second optical fiber to track the star light in the halo of the PSF. In this way, a contemporaneous spectrum of the star with tellurics can be captured and through cross-correlation techniques can be removed from the planet spectrum. 

The IRD instrument is fiber fed and has a slit consisting of $4$ fibers: $2$ single-mode fibers and $2$ multimode fibers. In order to maximize bandwidth and resolution, only one pair of fibers can be used at any one time. 

\begin{figure*}
\centering 
\includegraphics[width=0.90\linewidth]{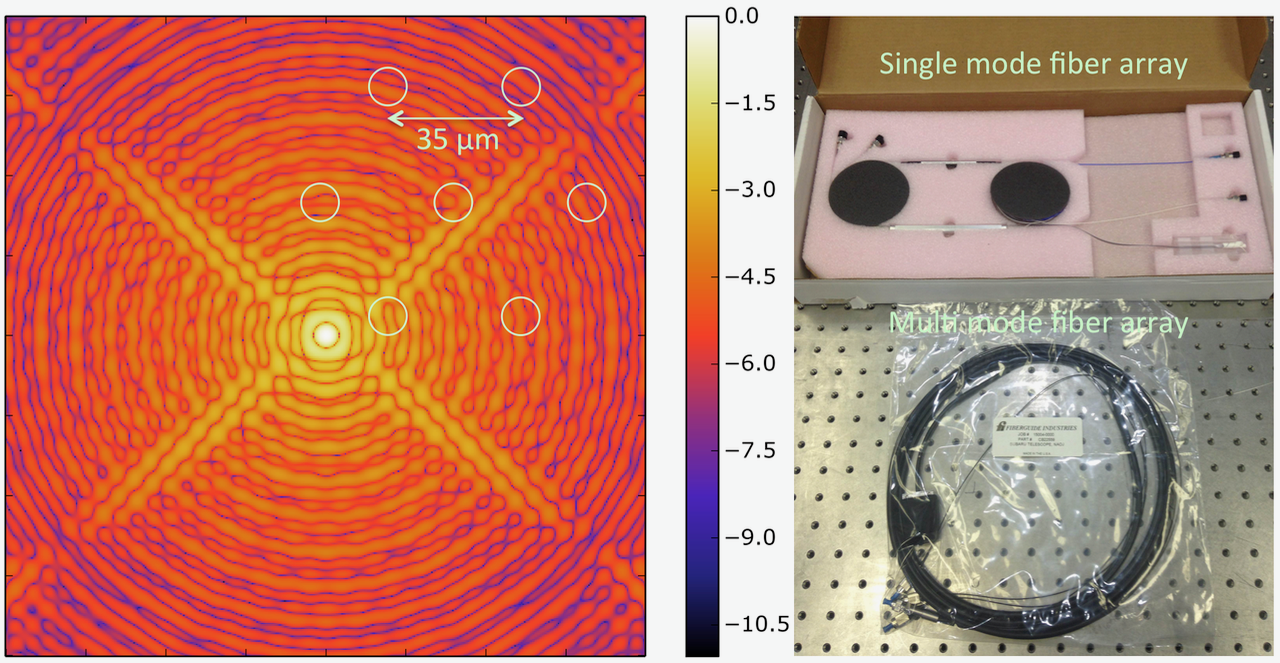}
\caption{\footnotesize (Left) Simulated diffraction limited PSF of SCExAO with an overlay of the positions and core size of the single-mode fiber array acquired for testing. (Right) An image of the two arrays.}
\label{fig:bundles}
\end{figure*}

Regardless of the fact that only $2$ fibers can be fed into the spectrograph at once we opted to use fiber bundles with $7$ fibers in a hexagonal geometry as shown in Fig.~\ref{fig:bundles}. The rationale for this is that the central fiber could be used to monitor the spectrum from the planet while the others, distributed throughout the speckle field of the star would track its spectrum. In this way, it would be possible to choose a fiber at a similar flux level to the planet fiber for monitoring the stellar spectrum, so that the planet fiber dictated the integration time used in IRD, not the other way around. We plan to develop an optical switch yard that will allow us to reconfigure which reference fiber is to be fed to IRD. 

The fiber injection has no vision system to indicate the location of the fibers in the field. This was chosen to minimize non-common optics and hardware in general. Instead, the bundles will be aligned in two stages. Firstly, the beam will be aligned on a particular pixel on the internal NIR camera during the day before going on-sky. The pickoff will be inserted and the light directed to the injection. The injection will be positioned to maximize the coupling into the fiber (i.e. on-axis). At this point the fiber will be considered co-aligned with the reference pixel of the internal NIR camera. Once on-sky, the star will be placed on this pixel before light is sent to the injection unit. The fiber will then need to be offset to the position of the known planet in the field. A specific fiber with the right flux levels will be chosen to monitor the stellar spectrum as outlined above. The other fibers could be used to track and keep the fiber aligned with the planet. This could be done by for example applying multiple sinusoidal modulations to the DM, which will generate speckles at the location of the 5 unused fibers in the focal plane. Then  a control loop could be used to maintain the flux coupled from the reference speckles into those $5$ fibers. Alternatively, a small raster scan could be implemented which scans each speckle across each fiber. In this way a two dimensional map of coupling versus spatial frequency could be computed which would identify any misalignments between the planet fiber and the planet itself. 

A more serious concern will be tracking the field rotation. SCExAO operates in fixed pupil mode in order to keep the corongraphic pupil masks aligned and the image in the PyWFS invariant. This means the field rotates. To track this the fiber injection unit will need to move in Cartesian coordinates to compensate. This compounds the difficulty of this goal as the planet is not visible itself and will be moving through the field so an accurate knowledge of the time is critical. 

\subsection{Speckle control}
Minimizing photon-noise will be critical to extracting high signal-to-noise ratio (SNR) planetary spectra. To achieve this we can go one step further and actively modulate the DM to enhance the rejection of the star light into the fiber. This is most easily achieved with the use of a single-mode fiber owing to the fact that light only needs to be rejected from one mode. Mawet et al. recently showed that suppressing quasi-static speckles could reduce the brightness in a single-mode fiber  by a factor of $100$--$1000$~\cite{mawet2017}. Even in the presence of turbulence a static solution to suppressing the quasi-static speckles resulted in an average reduction in the flux in the fiber by a factor of $\sim3$. We aim to use the commissioned speckle nulling algorithm on SCExAO~\cite{Nem2015a,martinache2014}, to suppress the light coupled from the quasi-static speckles into the fiber. To control the residual atmospheric speckles, a combination of predictive control and spatial linear dark field control (LDFC) will need to be implemented~\cite{guyon2017,miller2017}. The predictive control will be useful because the atmospheric speckles evolve rapidly (ms time scales) and this will allow the number of probes to be minimized allowing for more periods where the light can be integrated on for science. LDFC will be necessary as the dark hole, which by definition does not have much flux requires long integration times to sense the speckles. LDFC instead uses the bright speckles on the opposite side of the PSF to track small changes in the dark region so it can continue to run fast. For this application one of the other unused fibers in the bundle will be required. 

Implementing effective control loops to pursue and suppress atmospheric and quasi-static speckles will be challenging, but may offer significant contrast gains for HDC.

\subsection{The IR Doppler instrument}
IRD is a high resolution, $R\sim65000$ spectrograph that operates from y-H band~\cite{Kotani2014}. It is well stabilized in regards to both temperature and flexure and will be used for an RV survey around M-stars. The aim is to achieve a precision of $1$~m/s on brighter targets, enough to detect exo-Earths. This precision radial velocity machine is also ideal for characterizing exoplanet atmospheres as outlined above. 

IRD is currently undergoing commissioning and will begin routine science operation in early $2018$. This mode will then need to be commissioned soon after. This mode will offer the first high resolution ($R>10000$) spectra of exoplanet atmospheres to date along with contemporaneous spectra of the star and tellurics. We aim to use the stellar spectra to remove telluric contamination from the planet spectra and then cross-correlate it with the spectra of various molecular features to extract the low SNR signal from the data. New data reduction methods will need to be developed to take full advantage of the features of the fiber bundle (multiple fiber) and speckle modulation techniques outlined above. We hope to be in routine science operation from mid-2018.

\section{Outlook}
There are numerous challenging aspects to this project, but when successful this will offer a new approach to enhancing contrast to the level required for the characterization of exoplanets. SCExAO/IRD are an ideal combination that are nearly ready to begin the first experimental demonstration of HDC. This pioneering effort is being undertaken in parallel to the KPIC project at the Keck Telescope which aims to implement HDC with Keck AO and NIRSPEC for the characterization of known exoplanets in the K and L bands~\cite{wang2017,mawet2017}. Although HDC will be limited to large planets (Jovians) on $8$-m class ground based observatories, the larger collecting aperture and greater angular resolution of ELTs will open up the possibility for the characterization of terrestrial planets for the first time.

\acknowledgments 
 
The authors acknowledge support from the JSPS (Grant-in-Aid for Research \#$23340051$, \#$26220704$ \& \#$15$H$02063$). This work was supported by the Astrobiology Center (ABC) of the National Institutes of Natural Sciences, Japan and the directors contingency fund at Subaru Telescope. The authors wish to recognize and acknowledge the very significant cultural role and reverence that the summit of Maunakea has always had within the indigenous Hawaiian community. We are most fortunate to have the opportunity to conduct observations from this mountain. 

\bibliography{report} 
\bibliographystyle{spiebib} 

\end{document}